\documentclass[conference]{IEEEtran}
\IEEEoverridecommandlockouts
\usepackage{cite}
\usepackage{amsmath,amssymb,amsfonts}
\usepackage{algorithmic}
\usepackage{graphicx}
\usepackage{textcomp}
\usepackage[table]{xcolor}
\def\BibTeX{{\rm B\kern-.05em{\sc i\kern-.025em b}\kern-.08em
    T\kern-.1667em\lower.7ex\hbox{E}\kern-.125emX}}
\usepackage{orcidlink}    
\usepackage{booktabs,multirow,bigstrut}
\usepackage[nameinlink,noabbrev,capitalise]{cleveref}

\begin{document}

\title{Hybrid Event-Frame Neural Spike Detector for Neuromorphic Implantable BMI
\thanks{The work described in this paper was partially supported by a grant from Singapore Ministry of Education Academic Research Fund Tier 2 grant (MOE-T2EP20220-0002) and the Research Grants Council of the Hong Kong Special Administrative Region, China (Project No. CityU 11200922).}
}

\author{ 
Vivek~Mohan\textsuperscript{*}~\orcidlink{0000-0002-0248-6417}, Wee Peng Tay\textsuperscript{*}~\orcidlink{0000-0002-1543-195X} and Arindam~Basu\textsuperscript{\textdagger}~\orcidlink{0000-0003-1035-8770} \\ \textsuperscript{*}\textit{Nanyang Technological University, Singapore}~\textsuperscript{\textdagger}\textit{City University of Hong Kong}\\
Email: vivekmoh001@e.ntu.edu.sg, wptay@ntu.edu.sg, arinbasu@cityu.edu.hk
}

\maketitle

\begin{abstract}
This work introduces two novel neural spike detection schemes intended for use in next-generation neuromorphic brain-machine interfaces (iBMIs). The first, an Event-based Spike Detector (Ev-SPD) which examines the temporal neighborhood of a neural event for spike detection, is designed for in-vivo processing and offers high sensitivity and decent accuracy (94-97\%). The second, Neural Network-based Spike Detector (NN-SPD) which operates on hybrid temporal event frames, provides an off-implant solution using shallow neural networks with impressive detection accuracy (96-99\%) and minimal false detections. These methods are evaluated using a synthetic dataset with varying noise levels and validated through comparison with ground truth data. The results highlight their potential in next-gen neuromorphic iBMI systems and emphasize the need to explore this direction further to understand their resource-efficient and high-performance capabilities for practical iBMI settings.
\end{abstract}

\begin{IEEEkeywords}
implantable-brain machine interface (iBMI), neurotechnology, neuromorphic compression, event-based processing, spike detection
\end{IEEEkeywords}

\section{Introduction}
Significant developments in implantable brain-machine interface (iBMI) technology have transcended the initial purpose of studying the electrophysiological functioning of the brain, with successful demonstration of its potential to partially restore lost sensory capabilities such as vision and hearing through stimulation, and motor capabilities in people suffering from motor impairment or paralysis, in recent decades. While most existing iBMI systems comprise wired low-electrode count implants, next-generation implantable brain-machine interfaces intended for assistive technologies are expected to be wireless and capable of simultaneously sensing firing activity from thousands of neurons to enhance iBMI performance and functionality.

With Moore's law like scaling of the number of simultaneously recorded neurons by increasing the electrode count on the implant as studied in \cite{electrodeScaling}, digitization and transmission of neural data in the order of several gigabytes per second becomes a major hurdle, especially for a wireless implant, apart from limitations imposed by power dissipation and implant size. A neuromorphic compression based neural sensing pipeline (NCNS) inspired by asynchronous event-driven image sensors such as DVS \cite{dvs_tobi} was introduced in \cite{ISCAS2023}, which records neural activity as a stream of threshold crossing ON/OFF address-events, thereby ensuring $15-20\times$ and $2-5\times$ compression compared to conventional Nyquist rate full-sample and spike sample transmission respectively.

The extracellular aggregated electrical activity, captured by an electrode of an iBMI implant from the neurons in its vicinity, is called multi-unit activities (MUA). While the spiking activity of individual neurons called single-unite activity (SUA) can be isolated by spike sorting methods, it is computationally complex and found to be unnecessary for firing-rate based iBMI decoding\cite{naturePowerSaving}, which has demonstrated good performance with MUA alone \cite{Brain2Text, speechBMI, Todorova_2014}. Therefore, spike detection becomes an important problem for MUA-based iBMI, with several pioneering hardware architectures such as \cite{Rizk_2007}. Typical spike detection algorithms involve the determination of a noise threshold that is a scaled version of background noise determined from the band-pass filtered neural signal as done in the absolute negative threshold crossing method (AT-SPD)\cite{syntheticDataset} or its pre-emphasized form using a non-linear energy operator (NEO-SPD)\cite{NEO}, to separate the background activity from high amplitude spike signal.

Various implementations and approaches for spike detection have been proposed, ranging from hardware-efficient adaptive noise determination \cite{Valencia_2022} and firing-rate based threshold adaptation \cite{zzSPD,zz22, ZZ_FRM_SPD}, to machine learning based spike detection\cite{SpikeDeeptector, DL_SPD}. However, most of these works that show good spike detection performance are implemented assuming Nyquist rate conventional neural recording system whose resource requirements (power, bandwidth, memory, and size) might scale poorly with increasing electrode count \cite{ISCAS2023}. Spike detection in NCNS, which studies a neuromorphic iBMI system, requires recovery of the neural signal by stair-step reconstruction from the event stream as a precursor to non-linear energy operator (NEO) based spike detection on the recovered signal. \cite{Liu16} introduces an event-driven feature extraction for spike sorting, while \cite{NSS_PNS} presents a spiking neural network based temporal feature extractor, however, neither of these works explores event-based spike detection, keeping the subsequent spike decoding task in mind.

\begin{figure*}[t!]
\centering
\resizebox{0.9\textwidth}{!}{
\includegraphics[width=0.98\textwidth]{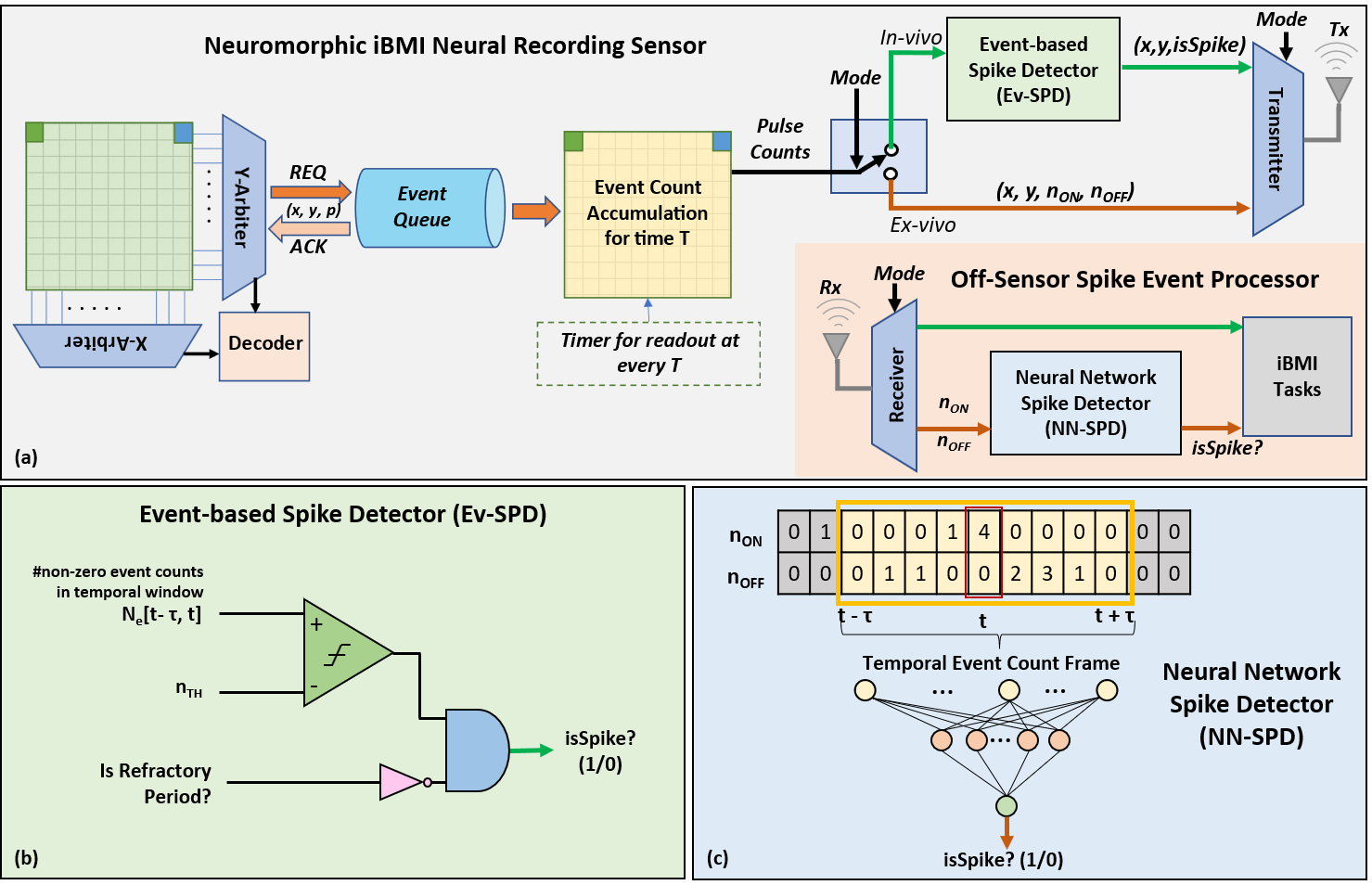}
}
\caption{(a) On-implant event-based spike detector (Ev-SPD) and Off-sensor neural-network spike detector (NN-SPD) detect spikes from the event counts generated by a neuromorphic iBMI neural recording sensor based on NCNS\cite{ISCAS2023}, operating in the pulse count mode (PCM). (b) The Ev-SPD consists of low-complexity operators that mark an event as a spike if there is a threshold number of non-zero PCM event counts in the temporal neighborhood of the current event. (c) The NN-SPD (MLP-SPD in this work) detects a spike from a hybrid temporal event frame centered around the spike peak.}
\label{fig:Figure1}
\end{figure*}

Taking advantage of background activity suppression in NCNS and leveraging on the compressed event data available in the pulse count mode (PCM), through this work, we try to fill the void in the area of event-based spike detection for the next generation of neuromorphic iBMI recording systems. We present two hybrid event-based spike detectors that can be used to detect spikes from the neural event data, in-vivo, and ex-vivo respectively, without additional signal recovery step. We make the following contributions:
\begin{itemize}
\item  We propose a low-complexity event-based spike detector (Ev-SPD) that takes advantage of the event generation pattern in NCNS and can be used as an on-implant coarse spike detection module for further the PCM data further, before wireless transmission for off-implant decoding.
\item We introduce shallow fully connected neural network based spike detectors (NN-SPD), specifically, multilayer perceptron (MLP-SPD) based spike detectors, that perform spike detection on a hybrid temporal event frame obtained from PCM event data and present a high-performance alternative to Ev-SPD suited for ex-vivo neuromorphic spike detection.
\end{itemize}

Here, instead of processing events in a conventional neuromorphic approach, we utilize the accumulated event counts per channel in PCM and process them along with event counts in a frame-like temporal window, resulting in a hybrid event-frame processing scheme for spike detection. 

\section{Methodology}
\subsection{Event Data Generation}
This work uses the event count data from the pulse count mode (PCM) introduced in NCNS \cite{ISCAS2023}. The pulse generation block in NCNS, as shown in Fig. \ref{fig:Figure1}, consists of a DVS-pixel-like ON and OFF threshold crossing detection circuit integrated into each neural recording cell. Considering a front end with a capacitive low-noise amplifier with gain $A_1$ and a programmable gain stage with gain $A_2$ (combined gain $\approx 200-1000$), the output $\mathrm{V_{mod}}$ and pulse $p\mathrm{(t)}$ generated by the delta modulator in the NCNS recording array is as follows:
\begin{gather}
\label{eq:pulse}
\frac{\mathrm{dV_{mod}}}{\mathrm{dt}}=A_1A_2\frac{\mathrm{dV_{in}}}{\mathrm{dt}}\notag\\
  p\mathrm{(t)} = 
\begin{cases} 
~~1,~\mathrm{V_{mod}(t+\delta)}=\mathrm{V_{cm}},~\text{if}~\mathrm{V_{mod}(t)}>\mathrm{Th_{ON}}, \\
-1,~\mathrm{V_{mod}(t+\delta)}=\mathrm{V_{cm}},~\text{if}~\mathrm{V_{mod}(t)}<\mathrm{Th_{OFF}}, \\
~~0,~\mathrm{V_{mod}(t+\delta)}=\mathrm{V_{mod}(t)}+\frac{\mathrm{dV_{mod}}}{\mathrm{dt}}\delta,~\text{otherwise,}
\end{cases}
\end{gather}
where, the input-referred ON/OFF pulse generation thresholds for a channel with mean spike amplitude of $\mathrm{V_{spike-max}}$ is obtained as:
\begin{align}
    \label{eq:thresholds}
 \mathrm{Th_{ON/OFF}}=\pm k \times \mathrm{V_{spike-max}}
\end{align}
In PCM, the ON/OFF events are accumulated in $\mathrm{n_{ON(OFF)}}$ event count bins of $\mathrm{t_{b}}$ duration, as follows:
\begin{flalign}
    \label{eq:pcm}
    \mathrm{n_{ON(OFF)}}=\sum_{\mathrm{t_{b}}} | p\mathrm{(t)}|_{p(t)=1(-1)}
\end{flalign}
\subsection{Event-based Spike Detection (Ev-SPD)}
The asynchronous delta modulator of NCNS generates events depending on the dynamics of the input signal. A fast rising or falling input signal such as spikes results in dense pulse generation, whereas a smooth/flat input signal results in few or no event pulses. Ev-SPD takes advantage of this characteristic of event generation to perform a low-complexity coarse spike detection suited for in-vivo implementation. As shown in Fig. \ref{fig:Figure1}(b), Ev-SPD detects the current event at time $\mathrm{t}$ as a spike, if the number of non-zero PCM event counts ($\mathrm{N_{e}}$) in the temporal neighborhood $\tau$ of the current PCM event exceeds a threshold of $\mathrm{n_{TH}}$, i.e., $\mathrm{isSpike=1}$ if $\mathrm{N_{e}}[\mathrm{t}-\tau, t]>\mathrm{n_{TH}}$. This ensures that only PCM events with sufficient temporal support are detected as a spike, while those due to background activity are filtered out. Once a spike is detected on a channel, the spike detector is disabled for a period determined by the refractory period $\mathrm{t_{ref}}$ (typically $>1$ ms) to prevent multiple detections per spike and spurious detections due to noise. When integrated with NCNS in the PCM mode, event data can be packetized as ($\mathrm{x,y,isSpike}$), resulting in further compression of the event data before transmission for downstream BMI tasks such as decoding. Assuming an average of $8$ events per spike, the combination of NCNS+Ev-SPD in-vivo can ensure $>8\times$ compression compared to NCNS event transmission for iBMI applications such as spike decoding where spike shape is not essential.

\subsection{Neural Network Spike Detector (NN-SPD)}
The shallow NN-SPD is a supervised machine learning trained spike detection algorithm, where the model is trained to discriminate the event count patterns around a spike from that of the background activity. For a non-zero event count $\mathrm{n_{ON/OFF}}$ at time $\mathrm{t}$, a hybrid temporal event-frame consisting of event counts in the neighborhood $\mathrm{t}\pm\tau$ around the present event is created and passed to the NN-SPD which then determines if the current event corresponds to the spike as shown in Fig. \ref{fig:Figure1}(c).
The NN-SPD models studied include - a variation of multilayer perceptron (MLP-SPD) with varying - depth (1 to 3 fully connected layers) and number of hidden neurons (16, 32, 64, and 128). The models were trained on balanced hybrid event-frames prepared from one of the single-channel recordings with varying noise levels, and the model with the best validation performance was saved for testing. The models were then tested using the remaining recordings in the dataset, in a continuous-time manner, with event-frames created dynamically whenever a non-zero event count is detected outside a refractory period. Similar to the implementation in Ev-SPD, a refractory filter disables the NN-SPD for a period $\mathrm{t_{ref}}$ on detecting a spike.

\section{Results}
\subsection{Dataset}
This work uses the popular synthetic dataset with varying noise levels of noise ($\mathrm{\sigma_{noise}}$ = 0.05, 0.1, 0.15, and 0.2) provided in \cite{syntheticDataset} along with the spike times that serve as `ground truth' for quantitatively measuring the performance of the spike detectors. Used widely to test spike detection performance in literature, this dataset simulates the neural activities of three neurons with an average firing rate of $20$Hz using the Poisson process. We use part of the Easy1 recording for training the NN-SPD models, and the Easy2 and Difficult1 recordings for testing the Ev-SPD and NN-SPD. The recordings sampled at $24$~KHz are converted to PCM events following the pipeline mentioned in \cite{ISCAS2023} with $k=0.2$ in Eq. (\ref{eq:thresholds}). The converted events are directly used for Ev-SPD, whereas temporal event frames were prepared beforehand for NN-SPD training and on the fly for testing.

\subsection{Evaluation Metrics} \label{evaluation_metrics}
Performance evaluation for the proposed spike detectors is done by comparing the detected spike times with the spike times provided in the ground truth of the dataset. Spike detections that lie within a window of $\mathrm{t_{spk}\pm\delta\mathrm{t}}$ ($\delta\mathrm{t}$ here is set to half of the average spike duration i.e., $0.5$~ms) are marked as true positives ($\mathrm{TP}$), spurious detections that are absent in the ground truth are marked as false positives ($\mathrm{FP}$) and the missed detections are marked as false negatives ($\mathrm{FN}$). Accuracy ($\mathrm{A}$), sensitivity ($\mathrm{S}$), and false detection rate ($\mathrm{FDR}$) calculated using Eq. (\ref{eq:metrics}) are used as metrics for measuring spike detection performance.
\begin{gather}
\label{eq:metrics}
\mathrm{S} =\frac{\mathrm{TP}}{\mathrm{TP + FN}};~ \mathrm{FDR} =\frac{\mathrm{FP}}{\mathrm{TP + FP}};~ 
\mathrm{A} =\frac{\mathrm{TP}}{\mathrm{TP + FP + FN}}
\end{gather}
\begin{table*}[t!]
\centering
\caption{Comparison of characteristics and implementations of different spike detection algorithms.}
\label{tab:sdp_comparison}
\resizebox{0.90\textwidth}{!}{%
\begin{tabular}{|l|c|c|c|c|c|}
\hline
\multicolumn{1}{|c|}{\textbf{Factor}} &
  \textbf{NEO/AT-SPD} &
  \textbf{Deep Learning SPD} &
  \textbf{Adaptive-Th SPD} &
  \textbf{Ev-SPD} &
  \textbf{Shallow NN-SPD} \\ \hline
\textbf{Work} &
  \cite{NEO,syntheticDataset} &
  \cite{SpikeDeeptector,DL_SPD} &
  \cite{zzSPD,zz22,Valencia_2022} &
  This work &
  This work \\ \hline
\textbf{Input Data} &
  \begin{tabular}[c]{@{}c@{}}Filtered/\\ pre-emphasized\\ neural signal\end{tabular} &
  \begin{tabular}[c]{@{}c@{}}Filtered data in\\ batches\end{tabular} &
  \begin{tabular}[c]{@{}c@{}}Filtered/\\ absolute-difference\\ neural   signal\end{tabular} &
  PCM Events &
  PCM Event Frame \\ \hline
\textbf{\begin{tabular}[c]{@{}l@{}}Calibration/ \\ Training Needed\end{tabular}} &
  Y &
  Y &
  N &
  N &
  \begin{tabular}[c]{@{}c@{}}Y (adaptable to different\\ noise levels)\end{tabular} \\ \hline
\textbf{Type} &
  Offline &
  Offline &
  On-implant &
  On-implant &
  Off-implant (wearable) \\ \hline
\textbf{Complexity} &
  Low &
  High &
  Low &
  Low &
  Medium \\ \hline
\textbf{Accuracy} &
  0.91-0.94 &
  0.95-0.97 &
  0.97 &
  0.94-0.97 &
  0.96-0.99 \\ \hline
\end{tabular}%
}
\end{table*}
\subsection{Ev-SPD Performance}

\begin{figure} [t]
\centering    
\includegraphics[width=0.48\textwidth]{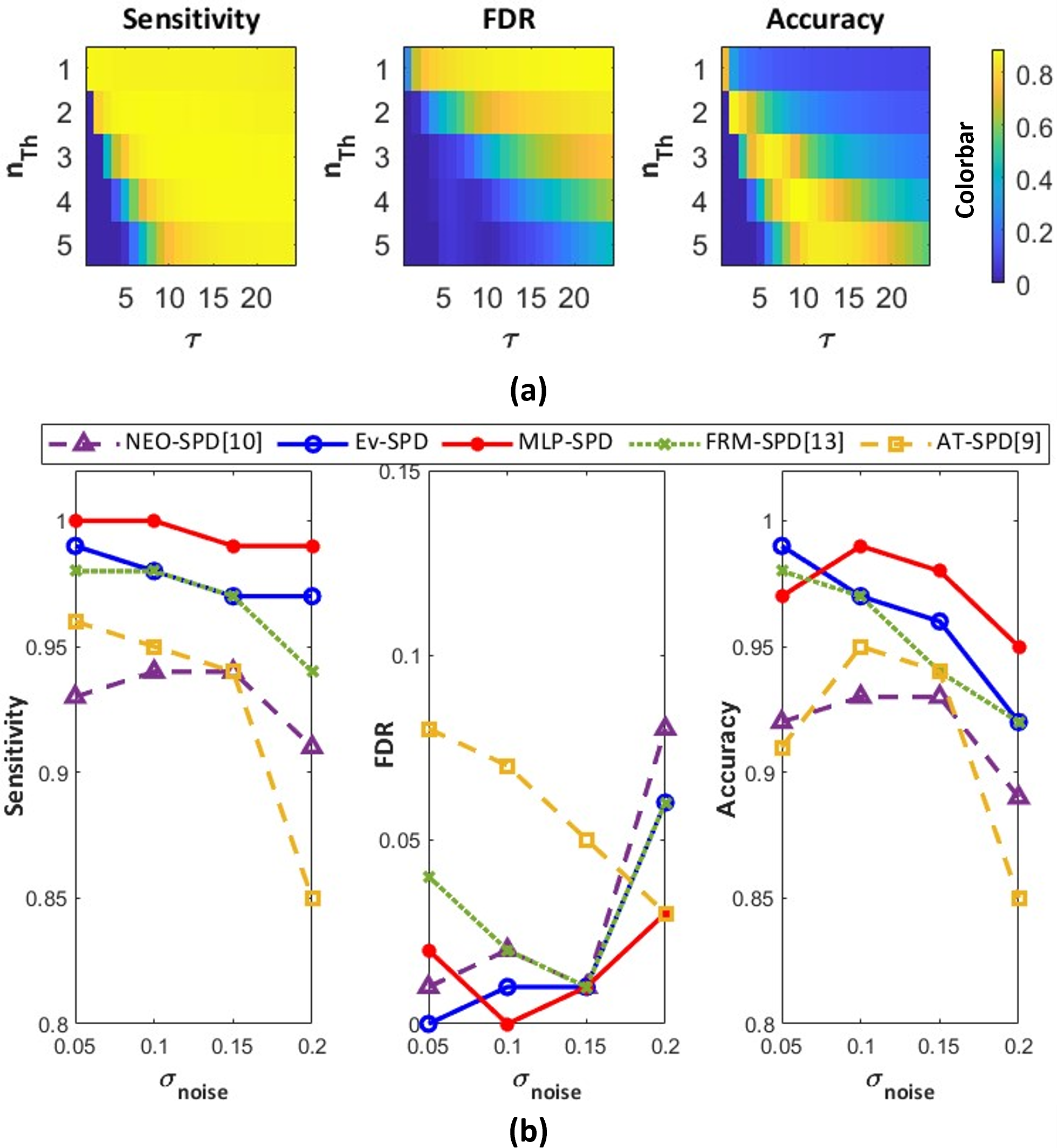}
 \caption{(a) Variation Ev-SPD performance with changes in event temporal neighborhood parameters $\mathrm{n_{Th}}$ and $\mathrm{\tau}$. The brighter regions in the heatmap indicate the higher values, and the darker regions correspond to the lower values of the computed metric. (b) Comparison of Ev-SPD and NN-SPD performance with other spike detection algorithms.}
\label{fig:Figure2}
\end{figure}

The performance of Ev-SPD is determined by the choice of the parameters $\mathrm{n_{TH}}$ and $\tau$. As shown in Fig. \ref{fig:Figure2}(a), a larger temporal window $\tau$ and smaller minimum non-zero PCM event count threshold $\mathrm{n_{TH}}$ for Ev-SPD results in higher sensitivity at the cost of increased false detections, whereas, a shorter $\tau$ and higher $\mathrm{n_{TH}}$ could risk missing spikes especially, those that are closer to the noise level. The accuracy of an Ev-SPD that requires support only from a PCM event in its immediate temporal neighborhood (i.e., $\mathrm{n_{TH}}=1~\text{and}~\tau=1$) varies between $0.72-0.98$. Despite its high sensitivity ($>0.98$), this variation results from an increased rate of false detections, particularly at higher noise levels, attributable to the lower threshold criteria. While this may not be ideal from a spike detection performance point of view, a higher sensitivity and slightly average FDR could ensure that smaller spike peaks closer to the noise floor are detected and in turn enhance the iBMI spike decoding performance bolstered by findings in \cite{Todorova_2014, Oby_2016, ZZ_FRM_SPD}. The results presented in Fig. \ref{fig:Figure2}(b) and Table \ref{tab:sdp_comparison} were obtained for $\mathrm{n_{TH}}=5$ and $\tau=11$ which approximately translates to a requirement of at least 4 non-zero PCM event counts in a half-spike wave temporal window of $0.5$~ms. As shown in Fig. \ref{fig:Figure2}(b), Ev-SPD performs relatively better than hardware-friendly firing rate based spike detectors (FRM-SPD) \cite{zz22} and the traditional NEO/AT-SPD. An on-implant Ev-SPD with a spike detection rate close to the biological neural firing rate ($\mathrm{f_{AP}=10-20}$~ Hz), spike-event-only transmission could result in a significant reduction in the NCNS data rates with over $8\times$ more compression compared to APM/PCM transmission in NCNS.

\subsection{NN-SPD Performance}
\begin{table}[t]
\centering
\caption{Variation of NN-SPD performance with neural network architecture}
\label{tab:NN-SPD}
\resizebox{0.48\textwidth}{!}{%
\begin{tabular}{|l|l|cc|ccc|}
\hline
\multicolumn{1}{|c|}{\multirow{2}{*}{\textbf{\begin{tabular}[c]{@{}c@{}}Input\\ Type\end{tabular}}}} &
  \multicolumn{1}{c|}{\multirow{2}{*}{\textbf{NN-SPD}}} &
  \multicolumn{2}{c|}{\textbf{Network Complexity}} &
  \multicolumn{3}{c|}{\textbf{Mean SPD Performance}} \\ \cline{3-7} 
\multicolumn{1}{|c|}{} &
  \multicolumn{1}{c|}{} &
  \multicolumn{1}{c|}{\textbf{\begin{tabular}[c]{@{}c@{}}MAC\\ (KOp/frame)\end{tabular}}} &
  \textbf{\begin{tabular}[c]{@{}c@{}}Memory\\ (Kb/frame)\\ (32-bit FP)\end{tabular}} &
  \multicolumn{1}{c|}{\textbf{S}} &
  \multicolumn{1}{c|}{\textbf{FDR}} &
  \textbf{A} \\ \hline
Filtered &
  2-FC MLP (32H) &
  \multicolumn{1}{c|}{12.3} &
  100.45 &
  \multicolumn{1}{c|}{0.997} &
  \multicolumn{1}{c|}{0.00065} &
  0.9965 \\ \hline
Recovered &
  2-FC MLP (32H) &
  \multicolumn{1}{c|}{12.3} &
  100.45 &
  \multicolumn{1}{c|}{0.9975} &
  \multicolumn{1}{c|}{0.00815} &
  0.9825 \\ \hline
\multirow{5}{*}{\begin{tabular}[c]{@{}l@{}}Event\\ Frame\\ (with sparsity \\ $\mathrm{S_{PCM}}\approx0.2$)\end{tabular}} &
  1-FC MLP &
  \multicolumn{1}{c|}{0.08} &
  0.62 &
  \multicolumn{1}{c|}{0.9975} &
  \multicolumn{1}{c|}{0.03305} &
  0.9655 \\ \cline{2-7} 
 &
  2-FC MLP   (16H) &
  \multicolumn{1}{c|}{1.35} &
  10.5 &
  \multicolumn{1}{c|}{0.9965} &
  \multicolumn{1}{c|}{0.027} &
  0.973 \\ \cline{2-7} 
 &
  \textbf{2-FC MLP   (32H)} &
  \multicolumn{1}{c|}{\textbf{2.7}} &
  \textbf{20.9} &
  \multicolumn{1}{c|}{\textbf{0.9975}} &
  \multicolumn{1}{c|}{\textbf{0.0195}} &
  \textbf{0.9795} \\ \cline{2-7} 
 &
  2-FC MLP   (64H) &
  \multicolumn{1}{c|}{5.4} &
  41.8 &
  \multicolumn{1}{c|}{0.999} &
  \multicolumn{1}{c|}{0.0085} &
  0.988 \\ \cline{2-7} 
 &
  3-FC MLP   (128H) &
  \multicolumn{1}{c|}{76.9} &
  620 &
  \multicolumn{1}{c|}{0.9985} &
  \multicolumn{1}{c|}{0.00055} &
  0.9985 \\ \hline
\end{tabular}%
}
\end{table}

MLP-SPD was trained and tested for the band-pass filtered neural recording and the signal recovered from NCNS events data as a baseline for comparison with NN-SPD performance on event frames. As shown in Table \ref{tab:NN-SPD}, the temporal event frame based NN-SPD performs well with significantly lower complexity - measured in terms of MAC operations and memory per frame, compared to filtered or recovered neural signals, owing to the sparsity of PCM events ($\mathrm{S_{PCM}}\approx20\%$). A 2-fully connected layer (FC) layer MLP-SPD with 32 hidden neurons was found to give a good trade-off between SPD performance and complexity and, was subsequently used for the remaining comparisons in this work. MLP-SPD consistently performs better than the other spike detection algorithms for varying noise levels, as shown in Fig. \ref{fig:Figure2}. While 32-bit floating point operation as performed on PyTorch was assumed for computation of memory, a quantized $4/8$-bit hardware implementation can significantly reduce the memory requirement, making them suitable for off-implant processing on wearables for downstream iBMI processing.

\section{Conclusion}
Neural spike detection remains an important problem and is the precursor for high-performance iBMI decoders. Addressing the absence of significant research in the direction of event-based spike detection for next-gen neuromorphic iBMI systems, this work proposes two spike detection schemes - the first a low-complexity event-based coarse spike detection scheme (Ev-SPD) which suitable for on-implant processing and event data compression, and the second a high-performance shallow neural network based spike detector (NN-SPD) for off-implant processing on devices such as wearables. Ev-SPD demonstrates high sensitivity and accuracy ($94-97\%$) better than or comparable to existing hardware-friendly spike detection schemes, whereas, NN-SPD consistently outperforms the other studied spike detectors with a mean spike detection accuracy of $96-99\%$ and negligible false detections. While results from this work present high-performance hybrid event-frame spike detectors based on analysis with the synthetic dataset, future works could analyze the implication of on-implant Ev-SPD and off-implant NN-SPD in terms of spike-detection performance for actual biological neural recordings and its effect on iBMI decoding performance.

\bibliographystyle{IEEEtran}
\bibliography{references}
\end{document}